\documentclass[10 pt,conference, compsocconf]{IEEEtran}
\IEEEoverridecommandlockouts
\newcommand{\subparagraph}{}
\usepackage{titlesec}
\usepackage{amsmath,amssymb,amsfonts}
\usepackage{wrapfig}
\usepackage{graphicx}
\usepackage{setspace}
\usepackage{tabularx}
\usepackage{float} 
\usepackage{multicol}
\usepackage{indentfirst}
\usepackage{cite} 
\usepackage{caption}
\usepackage[final]{pdfpages}
\usepackage{footnote}
\usepackage{textcomp}
\usepackage{algorithm}
\usepackage{algorithmic}
\usepackage{multirow}
\usepackage{url}
\usepackage[hang,flushmargin]{footmisc}
\usepackage[super]{nth}
\usepackage{titlesec}
\usepackage{subcaption}
\usepackage{fancyhdr}
\usepackage{bm}
\usepackage{newfloat}
\usepackage{epsfig}

\usepackage{enumitem}

\begin{document}
\title{CIDAN: Computing in DRAM with\\Artificial Neurons\vspace{-10pt}}


\author{\IEEEauthorblockN{Gian Singh, Ankit Wagle, Sarma Vrudhula\IEEEauthorrefmark{1}}
\IEEEauthorblockA{School of Computing and Augmented Intelligence\\
Arizona State University, Tempe, AZ, USA\\
(gsingh58, awagle1, vrudhula)@asu.edu}
\\[-4em]
\and
\IEEEauthorblockN{Sunil Khatri}
\IEEEauthorblockA{Dept. of Electrical and Computer Engineering\\
Texas A\&M University, College Station, TX, USA\\
sunilkhatri@tamu.edu}
\\[-4em]
\thanks{\IEEEauthorrefmark{1}This research was supported in part by NSF I/UCRC Center for Embedded Systems, NSF grant \#1361926.}
}		
\maketitle
\begin{abstract}
Numerous applications such as graph processing, cryptography, databases, bioinformatics, etc.,  involve the repeated evaluation of Boolean functions on large bit vectors.  In-memory architectures which perform \textit{processing in memory} (PIM) are tailored for such applications.  This paper describes a different architecture for in-memory computation called {\sc CIDAN}, that achieves a 3X improvement in performance and a 2X improvement in energy for a representative set of algorithms over the state-of-the-art in-memory architectures.  {\sc CIDAN} uses a new basic processing element called a TLPE, which comprises a threshold logic gate (TLG) (a.k.a \textit{artificial neuron} or \textit{perceptron}). The implementation of a TLG within a TLPE is equivalent to a \textit{multi-input, edge-triggered flipflop} that computes a subset of threshold functions of its inputs. The specific threshold function is selected on each cycle by enabling/disabling a subset of the weights associated with the threshold function, by using logic signals. In addition to the TLG, a TLPE realizes some non-threshold functions by a sequence of TLG evaluations. An equivalent CMOS implementation of a TLPE requires a substantially higher area and power. CIDAN has an array of TLPE(s) that is integrated with a DRAM, to allow fast evaluation of any one of its set of functions on large bit vectors. Results of running several common in-memory applications in graph processing and cryptography are presented. 
\end{abstract}

\section{Introduction}
The throughput of traditional Von Neumann architectures is severely limited by the bandwidth between CPU and memory due to the increasing gap in the performance of both the units. This well-known issue, often referred to as the "\textit{memory wall}", degrades the throughput of many high bandwidth applications that rely on bulk bitwise operations such as machine learning\cite{Angizi_He_Reis_Hu_Tsai_Lin_Fan_2019}, database management\cite{li_bitweaving:_2013}, encryption\cite{myers_fast_1999} etc. Over the last two decades, numerous researchers have proposed solutions to reduce the performance gap between the memory and the CPU. One category of solutions deals with improving the data bandwidth between memory and CPU. Examples of this category include double data rate (DDR) memory architectures and 3-D high bandwidth memory (HBM) architectures. While these solutions improve the throughput of the conventional memory, the greater improvement in CPU performance hasn't diminished the memory-wall significantly.  Another category of solutions involves increasing the size of the on-chip cache and introducing in-cache computations~\cite{fujiki_duality_2019}, \cite{eckert_neural_2018}. However, this method is limited by the amount of cache memory that can be added to the CPU chip. The third category of solutions is commonly known as processing-in-memory (PIM) architectures \cite{seshadri_ambit:_2017}, \cite{Newton_SK}, \cite{yin_vesti:_2020}. The idea here is to perform some computation within the memory and not involve the CPU in these computations. This solution has the potential to exploit the maximum parallelism of access inside the memory and can accelerate a plethora of applications. PIM architectures can reduce the number of data transactions on the channel connecting the CPU and the off-chip memory. Accordingly, PIM is one of the best-known methods to bypass the communication bottleneck, i.e., memory wall problem between CPU and memory. PIM offers unique opportunities to improve the parallelism and throughput of rapidly increasing highly parallel applications. Hence, PIM is an active and growing area of research.

In general, PIM architectures can be classified as mixed-signal PIM (mPIM) and digital PIM (dPIM) architectures. The dPIM architectures can be further classified as internal PIM (iPIM) and 
external PIM (ePIM). 


The mPIM architectures compute in the analog domain and convert the result into a digital value using an analog to digital converter (ADC). A few works representative of mPIM are \cite{yin_vesti:_2020}, \cite{PRIME_RRAM_2016}, \cite{Flash_array_2017}. mPIM architectures are based on either SRAM or non-volatile memories. They are extensively used for machine learning applications to perform multiply and accumulate (MAC) operations in parallel. mPIM architectures approximate the result as the accuracy depends on the precision of the ADC.


The iPIM architectures~\cite{seshadri_ambit:_2017},\cite{li_drisa:_2017}, \cite{deng_dracc:_2018}, \cite{angizi_redram:_2019} perform logic operations inside the memory array on one or two rows. iPIM architectures use internal DRAM operations with slight modifications to the memory array or the sense amplifier to achieve bitwise logic operation.

Due to the advancement in the memory technology to the 20nm node and the use 2.5D chip integration, a high bandwidth memory (HBM) has become possible. HBM uses bank-level parallelism to provide high bandwidth to the CPU. HBM has an even higher internal memory bandwidth which external PIM (ePIM) architectures exploit. The ePIM architectures embed digital logic outside the memory array but on the same die as the memory to use vast internal parallelism for machine learning applications. Recently DRAM makers SK-Hynix \cite{Newton_SK} and Samsung \cite{Samsung_FIM} introduce 16-bit floating-point processing units inside HBM. Though the existing PIM architectures can deliver much higher throughput as compared to the traditional CPU/GPU architectures\cite{angizi_redram:_2019}, they suffer from either loss of precision (mPIM architectures), low energy efficiency or high area overhead (ePIM architectures), or low throughput on complex operations (iPIM architectures). 

This paper presents a different design of a PIM architecture, named {\sc CIDAN}, which is a hybrid of the ePIM and iPIM designs.   {\sc CIDAN} introduces a new processing element, named \textit{threshold logic processing element} (TLPE), that comprises of a threshold logic gate (TLG) (a.k.a \textit{artificial neurons} or \textit{perceptron}).  A threshold function is a predicate involving a linear weighted arithmetic sum of its binary inputs.  The implementation of a TLG within a TLPE is equivalent to a \textit{multi-input, edge-triggered flipflop} that computes a \textit{subset} of threshold functions of its inputs. The specific threshold function is selected on each cycle by enabling/disabling a subset of the weights associated with the threshold function, by logic signals. In addition to the TLG, a TLPE realizes some non-threshold functions by a sequence of TLG evaluations. 
{\sc CIDAN} will be shown to substantially improve the energy efficiency and performance of state-of-the-art iPIM architectures.  The main contributions of the paper are listed below:
\begin{itemize}
\item This paper demonstrates a novel integration of threshold gates (artificial neurons) in a DRAM architecture to perform bulk bit-wise operations such as NOT, (N)AND, (N)OT, X(N)OR. This includes a novel structure for the threshold gate and its control circuit.

\item An extensive comparison against the state-of-the-art PIM architectures is presented for several applications such as data encryption, graph processing, and DNA sequence mapping, which benefits immensely from the PIM architectures. Demonstrations include 3X improvement in latency and energy over state-of-the-art architectures.
\end{itemize}

\section{Background}
\subsection{DRAM: Architecture, Operation and Timing Parameters}
\begin{figure}[h!]
\centering{\includegraphics[width=0.9\columnwidth]{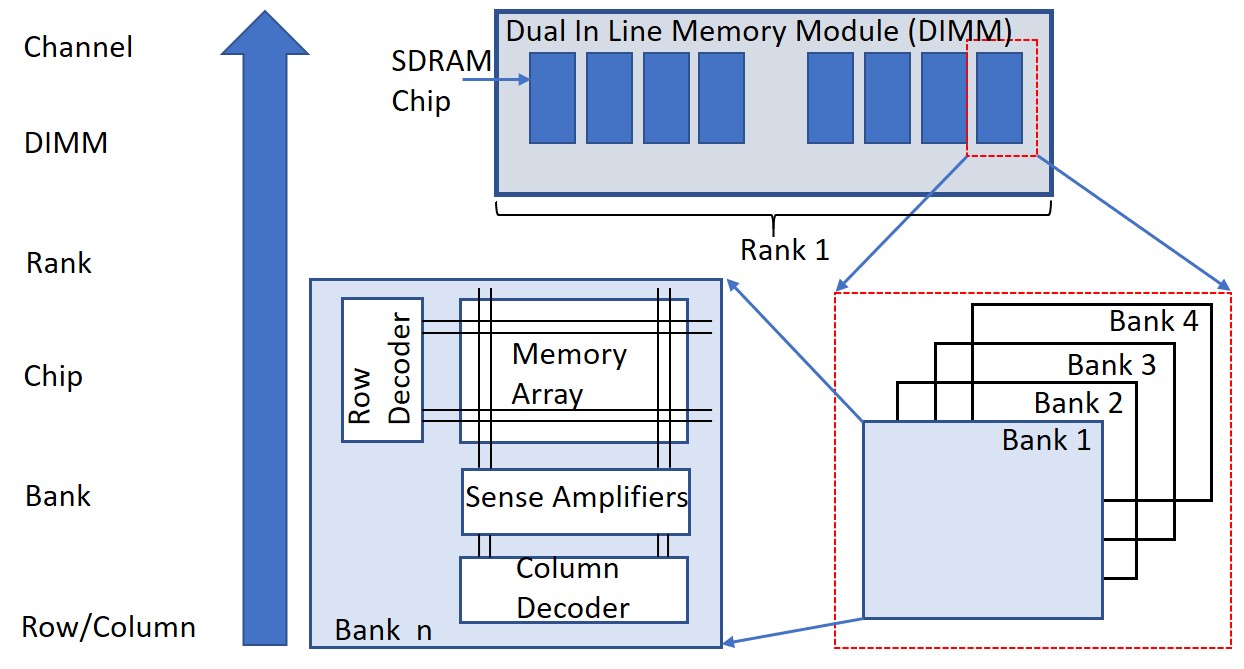}}
\caption{\small Top-level DRAM architecture.}
\label{fig:DRAM_arch}
\end{figure}

The organization of DRAM in modern computers is shown in Figure \ref{fig:DRAM_arch}. It consists of several levels of hierarchy. The lowest level of the hierarchy, which forms the building block of DRAM is called a \emph{bank}. A bank contains a 2D array of memory cells, a row of sense amplifiers, a row decoder, and a column decoder. A collection of banks is contained in a \emph{DRAM chip}. Memory banks in the chip share the I/O ports and an output buffer, and hence, only one bank per chip can be accessed at any given time. A group of DRAM chips is called a  \emph{Dual-in-line memory module (DIMM)}. Additionally, a \emph{rank} is a set of DRAM chips connected to the same chip select, which are therefore accessed simultaneously from a DIMM. When data needs to be fetched from the DRAM, the CPU communicates with the DRAM over a \emph{memory channel} with a data bus that is generally 64 bits wide. Multiple DIMM's can share a memory channel. Hence, a  multiplexer is used to select the DIMM to provide data to the CPU over the memory channel. One rank of the DIMM provides data to fill the entire data bus. Note that all the DRAM chips in a rank operate simultaneously while reading or writing data.

A DRAM memory controller controls the CPU access to the DRAM and the data transfers between the DRAM and the CPU. Therefore, almost all the currently available in-memory architectures modify the technique used to access the data and extend the functionality of memory to perform logic operations.  The controller issues a sequence of three commands to the DRAM: Activate (ACT), Read/Write (R/W), and Precharge (PRE), along with the memory address. The ACT command copies a row of data into the sense amplifiers through the corresponding bit-lines. Here, the array of sense amplifiers is called a \textit{row buffer} as it holds the data until another row is activated in the bank. The READ/WRITE command reads/writes a subset of row buffer to/from the data bus by using a column decoder. After the data is read or written, the PRE command charges the bit-lines to $\frac{VDD}{2}$, so that the memory is ready for the next operation. After issuing a command, the DRAM controller has to wait for an adequate amount of time before it can issue the next command. Such restrictions imposed on the timing of issuing commands are known as $timing$ $constraints$. Some of the key timing parameters are listed in Table~\ref{table:timing_constraints}.

\begin{table}[]
\small
\centering
\caption{ Important Timing Constraints of DRAM.}
\resizebox{0.8\columnwidth}{!}{\begin{tabular}{|c|c|}
\hline
\multirow{2}{*}{\textit{\textbf{t\textsubscript{RCD}}}} & time between the row access and \\ 
& data ready at Sense Amplifier \\ \hline
\multirow{2}{*}{\textit{\textbf{t\textsubscript{RAS}}}} & time between the access command \\ 
& and data restoration in DRAM array \\ \hline
\multirow{2}{*}{\textit{\textbf{t\textsubscript{RP}}}} & time taken to precharge DRAM array \\ 
& \\ \hline
\multirow{2}{*}{\textit{\textbf{t\textsubscript{RC}}}} & time interval between row accesses to \\ 
& two different rows in a bank t\textsubscript{RC} = t\textsubscript{RAS} + t\textsubscript{RP} \\ \hline
\multirow{2}{*}{\textit{\textbf{t\textsubscript{RRD}}}} & time interval between two row-activation \\ 
& commands to same DRAM device \\ \hline
\multirow{2}{*}{\textit{\textbf{t\textsubscript{FAW}}}} & time frame in which maximum four banks \\ 
& in DRAM device can be activated \\ \hline
\end{tabular}}
\label{table:timing_constraints}
\end{table}


Due to the power budget, traditional DRAM architectures allow only four banks in a DRAM to be activated simultaneously, within a time frame of $t_{FAW}$. The DRAM controller can issue two consecutive ACT commands to different banks separated by a time of $t_{RRD}$. As a reference, a 1Gb DDR3-1600 RAM has $t_{RRD} = 7.5ns$ and $t_{FAW} = 30ns$ \cite{DRAM_Power}. In Section~\ref{ARCH} the impact of these timing parameters on the delay in executing logic functions will be shown for the proposed processing-in-memory (PIM) architecture. 

\subsection{Prior Work on Logic Operations in DRAM (iPIM)}

\begin{figure}[h!]
\centering{\includegraphics[width=0.4\textwidth]{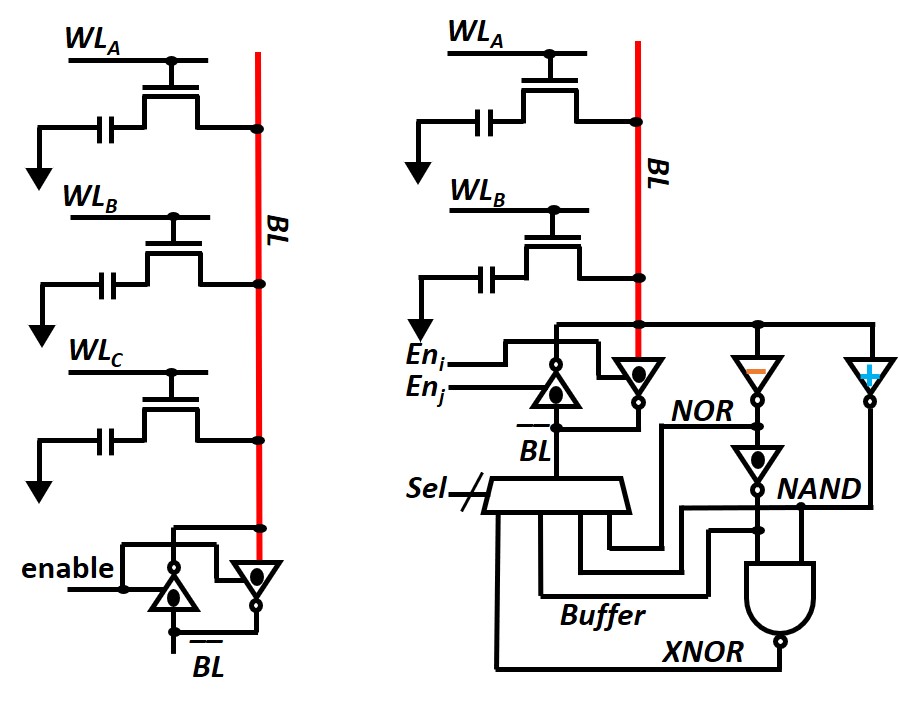}}
\caption{\small Hardware architecture of Ambit (Left) and ReDRAM (Right).}
\label{fig:prior_work}

\end{figure}

Currently available iPIM architectures such as Ambit \cite{seshadri_ambit:_2017} and ReDRAM \cite{angizi_redram:_2019} extend the operations of a standard DRAM to perform logic operations.
A representative diagram of Ambit operation on three rows is shown in Figure \ref{fig:prior_work} (left). Its operation is as follows: DRAM commands (ACT, PRE, etc.) are issued to fetch the operands, perform the logic operation, and then write back the result. It operates simultaneously on three rows which are activated using a special row decoder using triple row activation (TRA) operation. When three rows are activated simultaneously i.e., \textit{WL\textsubscript{A}, WL\textsubscript{B} and WL\textsubscript{C}} are turned ON, the capacitors are connected to the bit-line (BL). The charge in the capacitors used to store the values and the precharged \textit{BL} (at $\frac{VDD}{2}$) goes into the charge sharing phase until they stabilize to a voltage $\frac{VDD}{2} \pm \delta$. Using the resultant voltage at the end of the charge sharing phase, the sense amplifier computes the majority function by comparing this voltage against a reference and then driving \textit{BL} to logic 0 or 1. Since the rows remain connected to the \textit{BL}, the original data in the capacitors is lost and overwritten by the result value on \textit{BL}. 
In the case of ReDRAM, two rows are activated (double row activation-DRA) at a time and they undergo the same charge sharing phase with the \textit{BL} as in the case of Ambit.
To prevent the loss of original data at end of TRA or DRA, both Ambit and ReDRAM reserve some rows called \textit{compute rows} in the memory array to exclusively perform a logic operation. Hence, for every operation, the operands are copied from the source rows to the compute rows by using the copying operation described in the work Row Clone\cite{Row_clone}. A copy operation is carried out by a command sequence of $ACT\xrightarrow{}ACT\xrightarrow{}PRE$ which takes \textit{82.5ns} in 1Gb DDR3-1600 \cite{DRAM_Power}. In Ambit, all the 2-input operations such as AND, OR etc. are represented using a 3-input majority function. 


\begin{figure}[h!]
\vspace{-10pt}
\centering{\includegraphics[width=0.9\columnwidth]{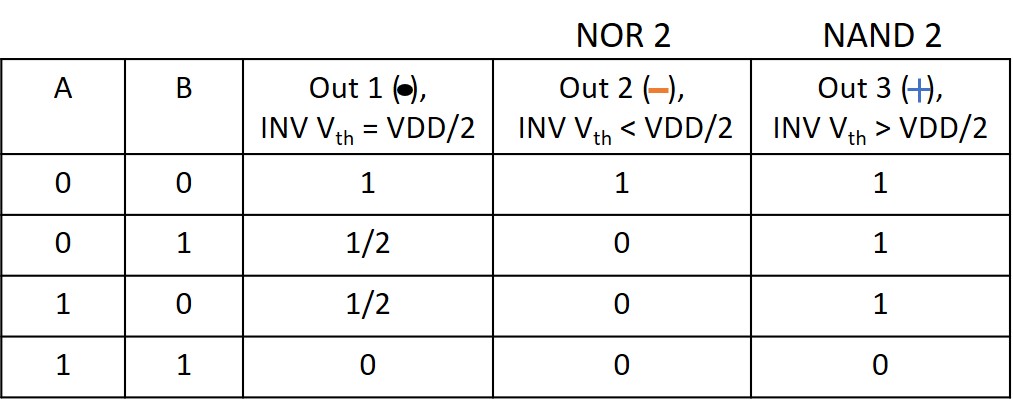}}
\caption{\small ReDRAM Inverters truth table for NOR2 and NAND2.\cite{angizi_redram:_2019}}
\label{fig:ReDRAM_op}
\end{figure}

ReDRAM improves upon the work by Ambit, by reducing the number of rows that need to operate simultaneously to two. After the charge sharing phase, but only between two rows, a modified sense amplifier is used to perform the logic operation and write back the result. The modified sense amplifier is used to change the reference voltage and perform the logic operations, as opposed to the sense amplifier in Ambit that only uses a fixed reference voltage. In ReDRAM, the reference voltages are changed with the help of inverters of varying operating points. Each inverter enables the computation of a different logic function as shown in the Figure \ref{fig:ReDRAM_op}. A multiplexer is used to choose between one of the logic functions.

ReDRAM and Ambit have a complete set of basic functions and can exploit full internal bank data width with a minimum area overhead. However, their shortcomings include: 
\begin{itemize}
\item Ambit and ReDRAM rely on sharing charges between the storage capacitors and bit-lines for their operation. Due to the analog nature of the operation, the reliability of the operation may get affected under varying operating conditions. 
\item ReDRAM modifies the inverters in the sense amplifier to shift their switching points using transistors of varying threshold voltage at the design time. Hence, such a structure is vulnerable to process variations. 
\item Both the designs overwrite the source operands, because of which rows need to be copied before performing the logic operations. Such an operation reduces the overall throughput that can be achieved when performing the logic operations on bulk data.
\end{itemize}

~\\\textbf{Key Advantages of CIDAN:} The proposed platform CIDAN improves the Ambit and ReDRAM iPIM architectures in four distinct ways. 
\begin{enumerate}
\item Memory bank and its access protocol are not modified.
\item CIDAN does not need special sense amplifiers for its operation.
\item Only a single row in a bank is activated. Multiple operands are fetched from different banks within the four bank activation window.  
\item  The TLPE(s) connected to the DRAM do not rely on charge sharing over multiple rows and are essentially \textit{static logic} gates. They are also reconfigured at run-time to realize different functions. 
\end{enumerate}
Further details are discussed in Section \ref{ARCH}.

\section{Processing-in-Memory Platform} \label{ARCH}
This section describes the architecture of {\sc CIDAN}.
\vspace{-3pt}
\subsection{Threshold Logic function and Artificial Neurons} \label{Background:TLG}
A Boolean function $f(x_1, x_2, \cdots, x_n)$ is called a threshold function if there exist weights $w_i$ for $i = 1, 2, \cdots, n$ and a threshold $T$\footnote{\vspace{-8pt} W.L.O.G. the weights $w_i$ and threshold $T$ can be integers \cite{book:muroga}.} such that
\begin{equation}
\label{eq:thresholddef}
f(x_1, x_2, \cdots x_n) = 1 ~\Leftrightarrow~ \sum_{i=1}^{n} w_i x_i \geq T,
\end{equation}
where $\sum$ denotes the arithmetic sum. Thus a threshold function can be represented as $(W, T) = [w_1, w_2, \cdots, w_n; T]$. An example of a threshold function is $f (a,b,c,d) = ab \lor ac \lor ad \lor bcd$, with $[w_1, w_2, w_3, w_4; T] = [2, 1, 1, 1; 3]$.
An extensive body of work exploring many theoretical and practical aspects of threshold logic can be found in \cite{book:muroga}. 

Several implementations of threshold logic gates already exist in the literature \cite{Jinghua_Fab_2015} \cite{Kulkarni_TVLSI_2016} \cite{FTL_2019} . These gates evaluate Equation \ref{eq:thresholddef} by directly comparing some electrical quantity such as charge, voltage, or current. For this paper, a variant of the architecture shown in \cite{FTL_2019} is used, as it is the threshold gate available at the smallest technology node (40nm).

\begin{figure}[h!]
\centering{\includegraphics[width=1\columnwidth]{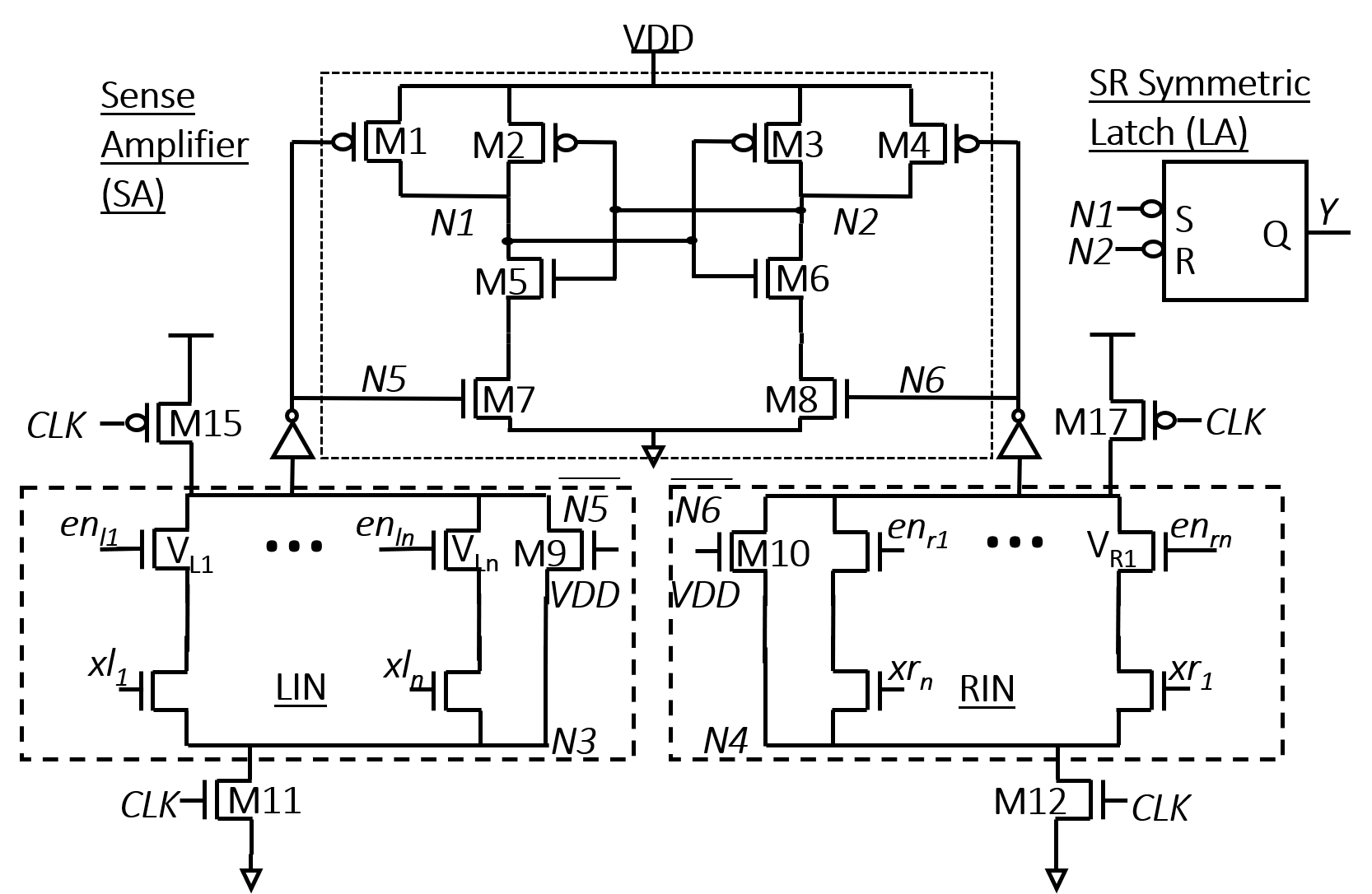}}
\caption{\small Threshold Logic Gate (TLG) Architecture.}
\label{fig:Generic_TLG}
\end{figure}

Figure~\ref{fig:Generic_TLG} shows the circuit diagram for the threshold gate. It consists of four components: left and right input network (LIN and RIN respectively), a sense amplifier, and a latch. When the clock signal is 0, the sense amplifier resets, nets $\overline{N5}$ and $\overline{N6}$ reset to 1 through transistor M15. This resets the sense amplifier through transistors M1 and M4 (N1=N2=1). For evaluation, appropriate input signals are provided to inputs $xl_1$ to $xl_n$ and $xr_1$ to $xr_n$, which in turn allow current to pass through the branches of LIN and RIN respectively. The current passing through the branches is proportional to the width of the transistors, which in turn serve as a proxy for the weights of the threshold function. Additional enable signals $en_{l1}$ to $en_{ln}$ and $en_{r1}$ to $en_{rn}$ have been incorporated to choose branches corresponding to the inputs that are being evaluated. During an evaluation phase, both the LIN and RIN discharge $\overline{N5}$ and $\overline{N6}$. Without the loss of generality, assuming $\overline{N5}$ discharges faster than $\overline{N6}$, M7 turns ON before M8, which enables the discharge of N1 faster than N2. N1 shuts off the transistor M6 and chokes the discharge path of N2. In the end, N1 is at 0 and N2 is at 1. The SR latch uses the differential output of the sense amplifier and evaluates to 1. Since the sense amplifier compares the conductivity of LIN and RIN, it serves as a proxy for the inequality shown in Equation \ref{eq:thresholddef}. LIN represents the left side of the equation and RIN represents the right side. Ensuring that the inputs to LIN and RIN are applied at a clock edge turns the circuit into a multi-input, edge-triggered flipflop, that computes the Boolean threshold function. Note that transistors M9 and M10 are added to prevent the $\overline{N5}$ and $\overline{N6}$ from any potential floating condition in case, all the branches are turned off.

\subsection{Threshold logic processing element (TLPE)}
\begin{figure}[h!]
\centering{\includegraphics[width=0.8\columnwidth]{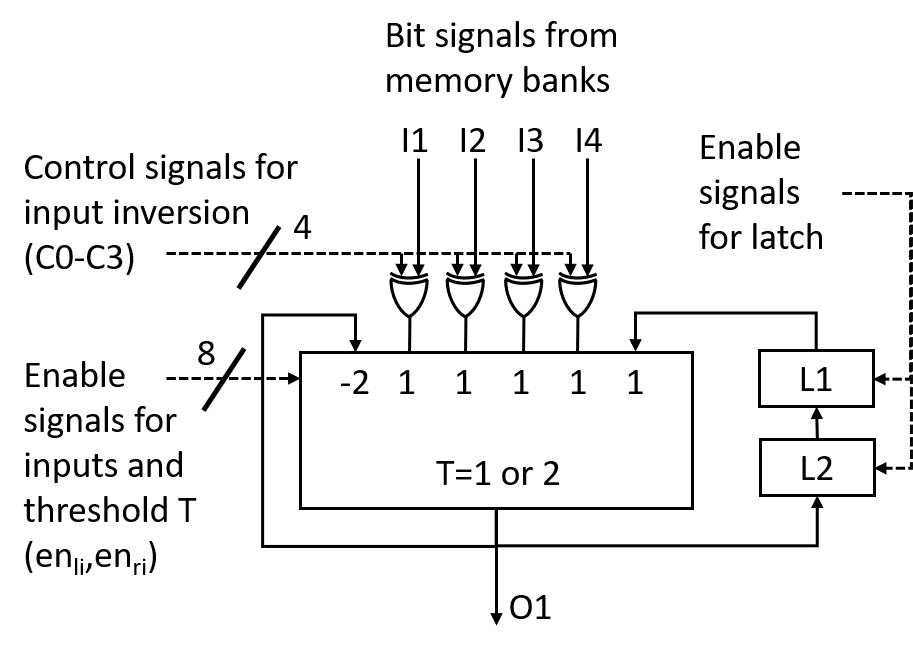}}
\caption{\small Architecture of threshold logic processing element.}
\label{fig:TLPE}
\end{figure}

Figure~\ref{fig:TLPE} shows the structure of a TLPE.  It consists of one threshold gate to perform computations, two latches L1 and L2 to temporarily store the output, and four XOR gates to invert the signals from the banks. The threshold gate is designed to implement a subset of the threshold function [-2,1,1,1,1,1;T], where T is selected during operation to be 1 or 2. The inputs to the processing element can be inverted using control signals \textit{C0-C3} or optionally disabled using enable signals \textit{en\textsubscript{li}}. The threshold and the remaining two inputs of the TLG are enabled or disabled by signals \textit{en\textsubscript{ri}}. All the control and enable signals are generated by an external controller discussed in section~\ref{sec:system-level_controller}. The TLPE is designed to perform basic logic operations (NOT, (N)OR, (N)AND, X(N)OR) and addition operation of 3 bits, which are the most commonly used operations for in-memory applications. 

Logic operations which are threshold functions, i.e., (N)AND and (N)OR functions are performed in a single cycle by enabling the required inputs (I1-I3) of the TLPE, and choosing an appropriate threshold T, as shown in Table \ref{table:basic_op}. Non-threshold functions such as XOR and XNOR operations are decomposed into two threshold functions and then scheduled on the TLPE over two cycles.  

\begin{figure}[h!]
\centering{\includegraphics[width=0.6\columnwidth]{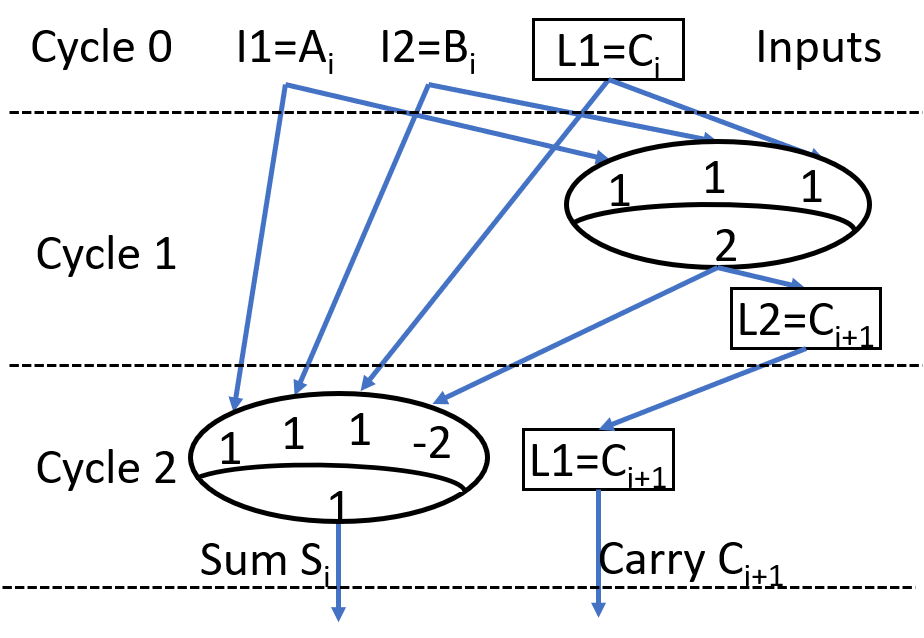}}
\caption{\small Schedule for addition operation on TLPE for 3-bits $A_i$, $B_i$ and $C_i$. Output is $S_i$ and $C_{i+1}$.}
\label{fig:adder_schedule}
\vspace{-10pt}
\end{figure}

The addition operation using a single threshold gate is slightly more complicated. It is a bit-level operation that is implemented on the TLPE in the form of a schedule, as shown in Figure \ref{fig:adder_schedule}. Assume that the $i^{th}$ significant bit of operands $A$ and $B$ are being added, with the previous carry $C[i]$ stored in L1. In the first cycle, carry $C[i+1]$ is generated by computing the majority function of the operand bits $A[i]$ and $B[i]$ and the previous carry $C[i]$. $C[i+1]$ is stored in the latch L2 and is also fed back as an input to the threshold gate. In the second cycle, sum-bit $S[i]$ is computed with the inputs $C[i+1]$, $A[i]$, $B[i]$ and $C[i]$ mapped to the  weighted threshold function [-2,1,1,1;1]. Once the sum bit is generated, the data in latch L2 is copied to latch L1, so that $C[i+1]$ is available for the adding next significant bits of operands $A[i+1]$ and $B[i+1]$.

\begin{table*}[]
\small
\centering
\caption{Power and area of functionally equivalent CMOS circuit  normalized to TLPE (TLPE=1). The columns labeled \textbf{M-Input-T} and \textbf{M-Input-X} are implementations of threshold functions and  XOR/XNOR functions for M= 2,3, and 4.}
\begin{tabular}{|l|c|c|c|c|c|c|c|c|c|c|}
\hline
                        & \multicolumn{7}{c|}{\textbf{Power}}                                                                                                          & \multicolumn{3}{c|}{\textbf{Area}}               \\ \hline
                        & \textbf{2-Input-T} & \textbf{3-Input-T} & \textbf{4-Input-T} & \textbf{2-Input-X} & \textbf{3-Input-X} & \textbf{4-Input-X} & \textbf{2-ADD} & \textbf{2-bit} & \textbf{3-bit} & \textbf{4-bit} \\ \hline
\multicolumn{1}{|c|}{\textbf{CMOS}} & 3.5x               & 4.3x               & 5.3x               & 1.7x               & 1.4x               & 1.3x               & 1.7x           & 1.9x           & 3.2x           & 4.0x           \\ \hline
\end{tabular}
\label{table:TLG_CMOS}
\end{table*}

\begin{table}[]
\small
\centering
\caption{Basic logic operations using threshold logic processing element. For demonstration, operands are I1 and I2.}
\begin{tabular}{|l|c|c|c|c|c|}
\hline
\multicolumn{1}{|c|}{\multirow{2}{*}{Function}} & \multirow{2}{*}{\begin{tabular}[c]{@{}c@{}}Cycle\\ number\end{tabular}} & \multicolumn{3}{c|}{Weights} & \multirow{2}{*}{T} \\ \cline{3-5}
\multicolumn{1}{|c|}{} & & -2 & 1 & 1 & \\ \hline
NOT & 1 & X & $\sim$I1 & X & 1 \\ \hline
AND & 1 & X & I1 & I2 & 2 \\ \hline
OR & 1 & X & I1 & I2 & 1 \\ \hline
NAND & 1 & X & $\sim$I1 & $\sim$I2 & 1 \\ \hline
NOR & 1 & X & $\sim$I1 & $\sim$I2 & 2 \\ \hline
XOR & 1 & X & I1 & $\sim$I2 & 2 \\ \hline
& 2 & OP1 & $\sim$I1 & I2 & 2 \\ \hline
XNOR & 1 & X & I1 & I2 & 2 \\ \hline
& 2 & OP1 & $\sim$I1 & $\sim$I2 & 2 \\ \hline
\end{tabular}
\label{table:basic_op}
\end{table}


\begin{figure}[h!]
\centering{\includegraphics[width=0.9\columnwidth]{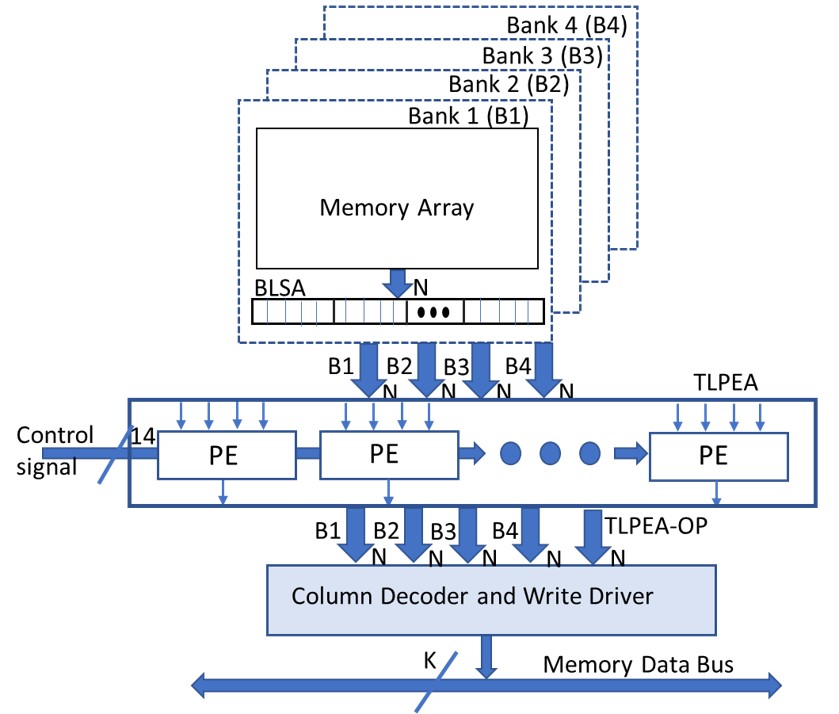}}
\caption{\small Threshold logic processing element array (TLPEA) connected to banks in a DRAM device.}
\label{fig:Bank_arch}
\vspace{-10pt}
\end{figure}

Note that it is possible to realize all the functionality of a TLPE using conventional CMOS logic gates. Table~\ref{table:TLG_CMOS} presents a comparison of a TLPE and its CMOS equivalent in terms of power and area. The CMOS implementation was obtained by jointly synthesizing all the functions including the selection logic. The columns labeled \textbf{M-Input-T} and \textbf{M-Input-X} are implementations of threshold functions and  XOR/XNOR functions for M= 2,3, and 4. \textbf{2-ADD} is the implementation of 2-bit adder. The results demonstrate that TLPE is significantly lower in area and power than the corresponding CMOS equivalent.  Note that an LUT could also be used, but this would be far worse than the custom CMOS implementation as it requires reprogramming on a cycle-by-cycle basis. 




\subsection{Top Level Architecture of CIDAN}  
Figure \ref{fig:Bank_arch} shows the integration of the threshold logic processing element (TLPE) within the DRAM memory chip. An array of TLPE (TLPEA) of size \textbf{N} is connected to four banks in the memory chip, where \textbf{N} is the number of bits in a row of the bank latched into the sense amplifiers (BLSA). There is one TLPEA for a set of \textbf{four} banks in one DRAM chip. A TLPEA accepts N-bit input vectors $ B1$, $B2$, $B3$ and $B4$ from all the four banks as shown in Figure \ref{fig:Bank_arch}. For bitwise operation in this work, at most two out of four banks are activated using a four-bank activation window ($t_{FAW}$) to get the operands. Consequently, only two out of four inputs to the TLPE are enabled by external control signals. The output of the TLPE array \emph{TLPEA-OP} is connected to the column decoder and write driver block as shown in the Figure \ref{fig:Bank_arch}. Using the control signals and the write drivers driven by the \emph{TLPEA-OP}, the result of computations is written back to one of the four banks. During the computation phase, the column decoder selects all the bit-lines of the selected bank. The data path of the PIM operations is orchestrated by the external controller described in section \ref{sec:system-level_controller}.

\subsection{System level integration and the Controller Design} \label{sec:system-level_controller}

CIDAN is used as a memory and as an external accelerator that is interfaced with the CPU. The design of CIDAN includes the addition of a special instruction to the CPU's instruction set that specifies the data and operation in CIDAN. Whenever the CPU identifies a CIDAN instruction, it passes it to the CIDAN controller~--~a state machine that decodes the CIDAN instruction and generates DRAM commands and control signals to implement the bitwise operation in CIDAN. Extra bits are added to the CPU-memory bus to accommodate the control signals from the CIDAN controller. Extending the instruction set and utilizing the DRAM controller have several advantages. \textit{First,} the operations in CIDAN can be directly triggered by the application instead of going through any other device API code which can cause additional overhead. \textit{Second,} CIDAN's memory can be accessed by the CPU directly which prevents copying data from CPU memory to the CIDAN memory as opposed to the memory of an FPGA. \textit{Finally,} the on-chip cache and the CIDAN memory can be kept coherent using the existing cache coherence protocols \cite{seshadri_ambit:_2017}. 

The CIDAN instruction is of the form, \textbf{bbop dest, src1, src2, func} where \textbf{bbop} specifies the CIDAN bulk bitwise operation, \textbf{dst} is the destination address, \textbf{src1} and \textbf{src2} are the source addresses and \textbf{func} is the logic function to be implemented in CIDAN. The instruction by default operates on a data of size equal to the CIDAN bank's row. For the data spanning multiple rows, the instruction must be repeated in the application code with different row addresses. If the data is less than the row size, it is assumed that dummy data is padded to complete the row.

The sequence of commands generated by the controller for all the CIDAN operations is shown in  Table \ref{table:cmd_seq}. An operation is carried on the data present in the row address $i$ and $j$. In CIDAN, $D_i$ and $D_j$ are read from bank $m$ and $n$ respectively and the result $D_r$ is written to bank $o$ row $r$. It should be noted that the consecutive activate commands in the case of CIDAN are issued to two different banks separated by $t_{RRD}$ time (7.5ns) and in the case of the other PIM platforms, the consecutive activate commands are issued to the same bank separated by $t_{RAS}$ time (35ns). It can be observed from Table \ref{table:cmd_seq} that while prior PIM platforms take multiple AAP (82.5ns each) commands to execute simple AND/OR functions and the number of such commands increases many times for complex functions like XOR whereas the command sequence for CIDAN remains short and constant. The PIM platforms: GraphiDe \cite{Graphide_Fan_2019} and SIMDRAM \cite{hajinazar_simdram:_2021} build upon ReDRAM and Ambit respectively perform addition and report (7 AAP) and (6 AAP + 2 AP) commands for 1-bit addition respectively. Hence, the advantage of using CIDAN increases for complex functions. 

\begin{table}[]
\small
\caption{ Basic Functions and Command Sequence for CIDAN and other PIM Platforms.\\$D_i$ = Data in row $i$, $A_{mi}$ = Activate bank $m$ row $i$, $W_{or}$ = Write bank $o$ row $r$, $PREA$ = Precharge all open banks, $AAP$ = ACT ACT PRE a bank, $AP$ = Activate Precharge a bank}
\resizebox{\columnwidth}{!}{\begin{tabular}{|c|c|c|c|c|c|}
\hline
\multirow{2}{*}{\textbf{Func}} & \multirow{2}{*}{\textbf{Operation}} & \multicolumn{4}{c|}{\textbf{Command Sequence}} \\ \cline{3-6} 
& & \textbf{CIDAN} & \textbf{ReDRAM} \cite{angizi_redram:_2019} & \textbf{Ambit} \cite{seshadri_ambit:_2017} & \textbf{DRISA}\cite{li_drisa:_2017} \\ \hline
\textbf{Copy} & \textit{D\textsubscript{r} $\xleftarrow{}$ D\textsubscript{i}} & \begin{tabular}[c]{@{}c@{}}  A\textsubscript{mi}\\ A\textsubscript{nr} \\1 clk cycle\\ W\textsubscript{nr}\\ PREA \end{tabular} & AAP & AAP & \begin{tabular}[c]{@{}c@{}}AP\\ AP\end{tabular} \\ \hline

\textbf{NOT} & \textit{D\textsubscript{r} $\xleftarrow{}$ $\overline{D\textsubscript{i} }$ } & \begin{tabular}[c]{@{}c@{}}  A\textsubscript{mi}\\ A\textsubscript{nr} \\1 clk cycle\\ W\textsubscript{nr}\\ PREA \end{tabular} & AAP & \begin{tabular}[c]{@{}c@{}}AAP\\ AAP\end{tabular} & \begin{tabular}[c]{@{}c@{}}AAP\\ AAP\end{tabular} \\ \hline

\textbf{AND} & \textit{ D\textsubscript{r}  $\xleftarrow{}$ D\textsubscript{i}  $\land$ D\textsubscript{j}  } & \begin{tabular}[c]{@{}c@{}} A\textsubscript{mi}\\ A\textsubscript{nj} \\ A\textsubscript{or} \\1 clk cycle\\ W\textsubscript{nr}\\ PREA \end{tabular} & \begin{tabular}[c]{@{}c@{}}AAP\\ AAP\\ AAP\end{tabular} & \begin{tabular}[c]{@{}c@{}}AAP\\ AAP\\ AAP\\ AAP\end{tabular} & \begin{tabular}[c]{@{}c@{}}AP\\ AAP\\ AAP\end{tabular} \\ \hline

\textbf{OR} & \textit{D\textsubscript{r}  $\xleftarrow{}$ D\textsubscript{i}  $\lor$ D\textsubscript{j}} & \begin{tabular}[c]{@{}c@{}}A\textsubscript{mi}\\ A\textsubscript{nj} \\ A\textsubscript{or} \\1 clk cycle\\ W\textsubscript{nr}\\ PREA \end{tabular} & \begin{tabular}[c]{@{}c@{}}AAP\\ AAP\\ AAP\end{tabular} & \begin{tabular}[c]{@{}c@{}}AAP\\ AAP\\ AAP\\ AAP\end{tabular} & N/A \\ \hline

\textbf{XOR} & \textit{D\textsubscript{r}  $\xleftarrow{}$ D\textsubscript{i}  $\bigoplus$ D\textsubscript{j}} & \begin{tabular}[c]{@{}c@{}}A\textsubscript{mi}\\ A\textsubscript{nj} \\ A\textsubscript{or} \\2 clk cycles\\ W\textsubscript{nr}\\ PREA\end{tabular} & \begin{tabular}[c]{@{}c@{}}AAP\\ AAP\\ AAP\end{tabular} & \begin{tabular}[c]{@{}c@{}}AAP\\ AAP\\ AAP\\ AP\\ AP\\ AAP\\ AAP\end{tabular} & N/A \\ \hline
\textbf{ADD} & \textit{\begin{tabular}[c]{@{}c@{}} D\textsubscript{r}  $\xleftarrow{}$ D\textsubscript{i}  $\bigoplus$ D\textsubscript{j}  $\bigoplus$ C\textsubscript{in}  \\ C\textsubscript{out} $\xleftarrow{}$ Maj(D\textsubscript{i}, D\textsubscript{j}, C\textsubscript{in}) \end{tabular}} & \begin{tabular}[c]{@{}c@{}}A\textsubscript{mi}\\ A\textsubscript{nj} \\ A\textsubscript{or} \\2 clk cycles\\ W\textsubscript{nr}\\ PREA\end{tabular} & N/A & N/A & N/A \\ \hline
\end{tabular}}
\label{table:cmd_seq}
\end{table}

\section{Experimental Results}
The performance and energy characteristics of CIDAN were evaluated using a combination of circuit-level and system-level simulations. The threshold gate used in the TLPE was evaluated for timing and proven for robustness using Monte Carlo simulations in TSMC 40nm LP technology. The TLPE was functionally verified using SPICE and its delay, energy, and area were extracted and scaled to the 45nm DRAM technology using \cite{DRAM_Logic_scaling_1996}. Gem5~\cite{Gem5_2011} was used for system-level simulation using these extracted values. Gem5 was integrated with Ramulator \cite{Ramulator_Mutlu_2016}~--~a DRAM simulator, to run the applications and obtain their performance statistics. The simulator DRAMPower~\cite{DRAM_Power} was used to evaluate the energy consumption. 

\subsection{Raw Performance and Energy}
CIDAN was evaluated and compared against the existing iPIM architectures, i.e., ReDRAM \cite{angizi_redram:_2019} and Ambit \cite{seshadri_ambit:_2017} for raw performance and energy. Evaluations were conducted on all the platforms using custom benchmarks that execute bulk-bitwise operations NOT, AND, OR, and XOR on large bit-vectors of size 1 Mb, 2 Mb, and 4 Mb. A uniform memory configuration of 8 banks with a memory array size of 16384x1024x8 bits was used. In Table \ref{table:op_latency_en} performance in terms of DRAM cycles required for each of the PIM platforms to execute each bitwise operation is presented. The data is normalized to the DRAM cycles taken by CIDAN.

\begin{table}[]
\small
\caption{Average Latency, Energy, and throughput for basic operations on PIM platforms computed for input vectors of size 1MB, 2MB and 4MB. Latency and energy is normalized to CIDAN.}
\resizebox{\columnwidth}{!}{\begin{tabular}{|c|c|c|c|c|c|c|c|}
\hline
                   & \multicolumn{2}{c|}{\textbf{Latency (CIDAN=1)}} & \multicolumn{2}{c|}{\textbf{Energy (CIDAN=1)}} & \multicolumn{3}{c|}{\textbf{Throughput (GOps/s)}} \\ \cline{2-8} 
\multirow{-2}{*} & \textbf{Ambit}         & \textbf{ReDRAM}        & \textbf{Ambit}        & \textbf{ReDRAM}        & \textbf{Ambit} & \textbf{ReDRAM} & \textbf{CIDAN} \\ \hline
\textbf{NOT}                                                      & 2.4                    & 1.2                    & 1.64                  & 0.82                   & 94.7           & 189.6           & 227.5          \\ \hline
\textbf{AND}                                                      & 4.32                   & 3.24                   & 2.61                  & 1.96                   & 47.3           & 63.1            & 205.03         \\ \hline
\textbf{OR}                                                       & 4.32                   & 3.24                   & 2.61                  & 1.96                   & 47.3           & 63.1            & 205.03         \\ \hline
\textbf{XOR}                                                      & 6.54                   & 3.19                   & 4.12                  & 1.94                   & 30.7           & 63.1            & 201.8          \\ \hline
\end{tabular}}
\label{table:op_latency_en}
\vspace{-10pt}
\end{table}

~\\\noindent \textbf{Performance:} As shown in Table \ref{table:op_latency_en}, ReDRAM requires about 3X more DRAM cycles than CIDAN to compute bitwise AND, OR, and XOR for different operand sizes. These improvements stem from the fact that CIDAN required far less internal DRAM operations than ReDRAM and any other PIM platform. CIDAN takes advantage of the four-bank activation window to activate two rows in different banks and has its operand ready in $t_{RRD} + t_{RCD}$ time (22.5ns). In contrast, the ReDRAM and other PIM platforms must perform a series of row-initialization and row copy operations using AAP (82.5ns) to get the operands ready for computation. CIDAN performs the function in one or two clock cycles inside the TLPE and then writes the result to another bank. Precharge in CIDAN is shared by all open banks and helps to decrease the latency. Whereas in ReDRAM and other PIM platforms, due to the series of ACT ACT PRE (AAP) operations, several clock cycles are needed to perform the operation and write the result.

~\\\noindent \textbf{Energy:} Table \ref{table:op_latency_en} compares the energy needed for primitive operations on ReDRAM, Ambit and CIDAN. On average, CIDAN's energy consumption is nearly half the energy of ReDRAM for bulk bitwise operations, and is significantly better than Ambit. Note that the values in Table \ref{table:op_latency_en} account for the fact that CIDAN activates more than one bank at a time, while ReDRAM and Ambit activate a single bank.

\subsection{Area Overhead}
The array of threshold processing elements (TLPEA) is added to the on-pitch area of the DRAM. Each TLPEA contains as many TLPE(s) as the number of bits in a memory bank's row. One TLPEA is shared by four banks within a DRAM chip. Note that the area of one TLPE is 41 
$\mu m\textsuperscript{2}$ as shown in the Table \ref{table:TLG_CMOS}. Therefore, for the DRAM configuration of 8 banks with 16384 rows, 1024x8 columns, the TLPEA occupies less than $\sim 1\%$ of the DRAM chip area.

\section{ Application-specific analysis: Case study}
The PIM platform CIDAN can accelerate the applications that make extensive use of bulk bit-wise operations on large bit vectors. Such applications are common in database management, graph processing, encryption, web search, bio-informatics, etc. In the following section, three algorithms have been chosen to demonstrate the performance and energy consumption of CIDAN.  They are the Advanced Encryption Standard (AES), graph processing operations, and DNA sequence mapping. In these applications, most of the operations are bitwise operations on large vectors. We expect the improvements of CIDAN over ReDRAM and Ambit in these applications to be proportional to the number and ratio of different bitwise operations used in these applications. 

\subsection{Advance Encryption Standard (AES)}

\begin{figure}[h!]
\centering{\includegraphics[width=1\columnwidth]{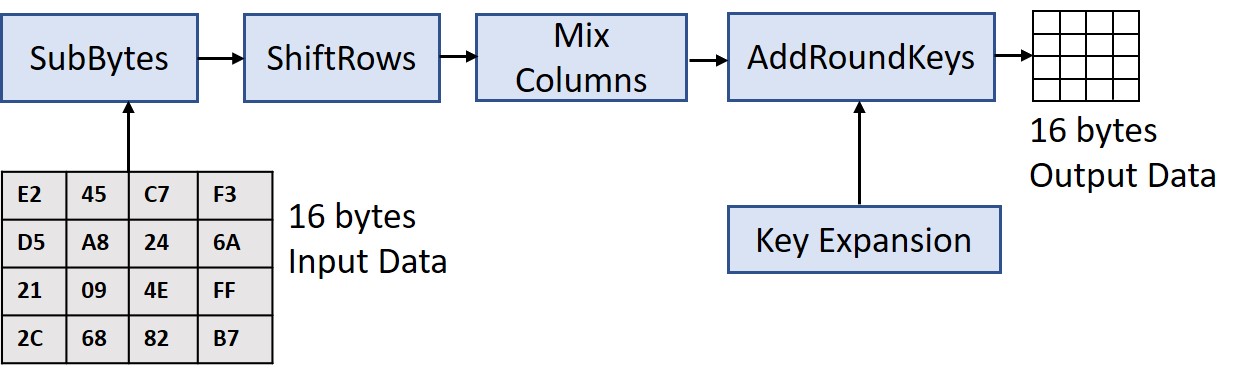}}
\caption{\small AES algorithm and operations details at every stage.}
\label{fig:AES_algo}
\end{figure}

AES is a block cipher standard that encrypts and decrypts data with a unique key. It takes 16 bytes (128 bits) of data in a 4x4 matrix form as input and uses three different key lengths (128 bits, 192 bits, and 256 bits) to encrypt the data through a series of transformations using bitwise operations. Various stages in the AES algorithm are described in Figure \ref{fig:AES_algo}. The encryption process takes place in multiple rounds of data transformation through stages shown in Figure \ref{fig:AES_algo}. If the key length is equal to 128 bits, then there are 10 rounds, 12 rounds for 192 bits key, and 14 rounds for 256 bits key.

To accelerate the encryption process, the stages Mix Columns and Add Round keys are executed on the PIM platforms. These two stages consist of mainly bitwise XOR and AND operations. Add round keys and Mix column stages account for about 75\% of the workload and we evaluate CIDAN, ReDRAM, and CPU with configuration as shown in Table \ref{table:CPU}.
\begin{table}[]
\small
\caption{ CPU configuration used in the experiments.}
\resizebox{\columnwidth}{!}{\begin{tabular}{|c|c|}
\hline
\multirow{2}{*}{Processor} & x86, 4 cores, out-of-order, 2 Ghz \\ \cline{2-2} 
& 64-entry instruction queue \\ \hline
L1 Cache & 32 KB D-cache, 32 KB I-cache, LRU policy \\ \hline
L2 cache & 256 KB, LRU policy, 64 B Cache line size \\ \hline
Main Memory & DDR3-1600, 1 channel, 1 rank, 8 banks \\ \hline
\end{tabular}}
\label{table:CPU}
\vspace{-10pt}
\end{table}

The performance in terms of total clock cycles required for CPU, ReDRAM, and CIDAN are given in Table \ref{table:AES_perf}, which shows that CIDAN achieves higher performance than ReDRAM and Ambit. Substantial improvements are gained from executing AES on CIDAN than on a CPU. The 75\% of the workload which was offloaded to CIDAN is reduced by ~40X. Similarly, the energy consumption of CIDAN is lowest among all the platforms and improves substantially over CPU.

\begin{table}[]

\small
\caption{ Latency and Energy comparison for executing AES on different platforms normalized to CIDAN.}
\resizebox{\columnwidth}{!}{\begin{tabular}{|l|c|c|}
\hline
\textbf{} & \textbf{Latency (CIDAN = 1)} & \textbf{Energy (CIDAN = 1)} \\ \hline
\textbf{ReDRAM} & 1.15 & 1.10 \\ \hline
\textbf{CPU} & 4.04 & 3.74 \\ \hline
\end{tabular}}
\label{table:AES_perf}

\end{table}

\subsection{Graph Analysis}
Graphs are traditional data structures that can store data value and also represent the relationship among data. Graphs have been a popular choice to build social networks. Facebook, Google+, and many other social networks use graphs to maintain the information of their users. With the rapid growth in the size of social networks, graph processing has become difficult and slow on the traditional Von-Neumann architectures. For several graph operations, PIM architectures are better suited. One such graph processing application is called \textit{matching index}.

Matching Index \textit{M\textsubscript{i,j}} is the similarity between the two vertices $V_i$ and $V_j$ based upon the number of common neighbors shared by these vertices and is given by
\[
\frac{\sum common\_neighbours}{\sum total\_number\_of\_neighbours}.
\]
To perform a matching index operation in memory, graph partitioning and allocation on memory is required. We use METIS\cite{Metis_1998} algorithm to do balanced graph partitioning. 

After mapping a graph in the form of adjacency matrices to CIDAN banks, the calculation of the number of common neighbors between two vertices $V_i$ and $V_j$ is just a bitwise AND operation between the two rows of $V_i$ and $V_j$ and the total number of neighbors is given by OR operation on the same two rows. The summation operation henceforth can be carried out in the CPU. 

To evaluate CIDAN and other iPIM platforms, a memory bank configuration of 16384 rows and 1024 columns of 8 bits each is used. There are in total 8 memory banks which makes the total capacity of the memory to be 128 MB. We evaluate three Social Network data sets as shown in Table \ref{table:social_nw}. 

\begin{table}[]
\small
\caption{ Social Network datasets.}
\resizebox{\columnwidth}{!}{\begin{tabular}{|c|c|c|c|}
\hline
\textbf{Dataset} & \textbf{Nodes} & \textbf{Edges} & \textbf{Graph Information} \\ \hline
\textbf{Facebook} & 4,039 & 88,234 & Social circles from Facebook \\ \hline
\textbf{DBLP} & 317,080 & 1,049,866 &  DBLP collaboration network \\ \hline
\textbf{Amazon} & 334,863 & 925,872 & Amazon product network \\ \hline
\end{tabular}}
\label{table:social_nw}
\vspace{-5pt}
\end{table}

In Table \ref{table:Graph_perf}, the relative performance of the iPIM architectures for matching index operation on two vertices relative to CIDAN is shown. Based on the results, CIDAN is the fastest architecture because of the computation on the TLPEA and the elimination of several internal DRAM operations. CIDAN also achieves the highest energy efficiency as shown in Table \ref{table:Graph_perf}. The non-PIM architectures like CPU and GPU as shown in \cite{angizi_redram:_2019} consumes 21X (GPU) more energy  than ReDRAM due to a large number of data transfers between computer unit and memory. It is also seen that both CPU and GPU spend most of the time waiting for the data to arrive in the compute unit to perform the operation. 

\begin{table}[]
\small
\caption{ Latency and Energy comparison for executing Matching Index operation on different platforms normalized to CIDAN.}
\resizebox{\columnwidth}{!}{\begin{tabular}{|c|c|c|c|c|c|c|}
\hline
\textbf{} & \multicolumn{3}{c|}{\textbf{Latency (CIDAN = 1)}} & \multicolumn{3}{c|}{\textbf{Energy (CIDAN = 1)}} \\ \hline
\textit{\textbf{Dataset}} & \textbf{facebook} & \textbf{amazon} & \textbf{dblp} & \textbf{facebook} & \textbf{amazon} & \textbf{dblp} \\ \hline
\textbf{ReDRAM} & 3.24 & 3.24 & 3.24 & 1.96 & 1.96 & 1.96 \\ \hline
\textbf{Ambit} & 4.32 & 4.32 & 4.32 & 2.61 & 2.61 & 2.61 \\ \hline
\end{tabular}}
\label{table:Graph_perf}
\vspace{-5pt}
\end{table}

\subsection{DNA sequence mapping}
DNA sequence mapping is an important application in bio-informatics which involves the pattern matching problem of a given DNA sequence to a large reference genome sequence. This problem can be solved as a string-matching problem by using the bit-vector algorithm\cite{myers_fast_1999} involving a large number of bitwise operations. Hence, such algorithms can easily be executed on PIM platforms for high throughput and energy efficiency. We implement and compare algorithm in \cite{myers_fast_1999} on CIDAN, ReDRAM and Ambit as shown in the Table \ref{table:DNA_perf}. The numbers in the table are normalized to that of CIDAN. ReDRAM and Ambit take about 3.14X and 4.35X more cycles respectively to execute the workload and consumes about 2.12X and 2.88X more energy, respectively.

\begin{table}[]
\vspace{-10pt}
\small
\caption{ Latency and Energy comparison for executing DNA sequence mapping algorithm\cite{myers_fast_1999} on different platforms normalized to CIDAN.}
\resizebox{\columnwidth}{!}{\begin{tabular}{|l|c|c|}
\hline
\textbf{} & \textbf{Latency (CIDAN = 1)} & \textbf{Energy (CIDAN = 1)} \\ \hline
\textbf{ReDRAM} & 3.14 & 2.12 \\ \hline
\textbf{Ambit} & 4.35 & 2.88 \\ \hline
\end{tabular}}
\label{table:DNA_perf}
\vspace{-10pt}
\end{table}
\section{Conclusion}
In this paper, a novel processing-in-memory (PIM) architecture is presented, which uses highly reconfigurable and low-power threshold logic processing elements (TLPE). By using these elements, basic two-input bit-wise logic functions such as NOT, (N)AND, (N)OR, X(N)OR, etc. and full adder operation can be accelerated inside a DRAM to achieve higher performance than equivalent state-of-the-art architectures. CIDAN takes advantage of the four bank activation window and TLPE(s) to compute bit-wise logic operations in about 3x less time than the state-of-the-art ReDRAM. The simulation results for the AES encryption, graph processing operation, and DNA sequence mapping algorithm shows superior performance and energy over any other iPIM architecture.

\nocite{TULIP}
\bibliography{references}
\end{document}